\documentstyle[multicol,aps,epsf]{revtex}

\title{{\mbox{\boldmath $XY$}} Spin Fluid in an External Magnetic Field}

\author{I. P. Omelyan,$^{1,2}$ W. Fenz,$^2$ I. M. Mryglod,$^{1,2}$
        and R. Folk,$^2$ }

\address{$^1$Institute for Condensed Matter Physics,
         1 Svientsitskii Street, UA-79011 Lviv, Ukraine}
\address{$^2$Institute for Theoretical Physics, Linz University,
         A-4040 Linz, Austria}

\date{\today}

\begin{document}

\maketitle

\begin{abstract}

A method of integral equations is developed to study inhomogeneous
fluids with planar spins in an external field. As a result, the
calculations for these systems appear to be no more difficult than
those for ordinary homogeneous liquids. The approach proposed is
applied to the ferromagnetic $XY$ spin fluid in a magnetic field
using a soft mean spherical closure and the Born-Green-Yvon
equation. This provides an accurate reproduction of the
complicated phase diagram behavior obtained by cumbersome
Gibbs ensemble simulation and multiple histogram reweighting
techniques.

\vspace{6pt}

\noindent
PACS number(s): 05.70.Fh, 64.60.-i, 64.70.Fx, 75.50.Mm

\end{abstract}

\vspace{0pt}

\begin{multicols}{2}

Spin fluids are examples of many body systems showing a {\em rich}
variety of phases in the global phase diagram \cite{HemImb77,%
TaGaTeWeNi95,Schinagl,Fefomrom03}. Besides gas-liquid (G-L) and
para-ferro-magnetic (P-F) phase transitions, tricritical, critical
end and triple point behavior is observed. Under special conditions,
an unsymmetrical tricritical van Laar point exists additionally
\cite{OmMrFoFe04}. This complexity arises due to a {\em coupling}
between spin and spatial interactions. Similar phase diagrams
are found in symmetric binary mixtures \cite{Wilding,Kahlang,%
Kahl,Paschinger,Pini} with their demixing and G-L transitions,
spin lattice gas models \cite{Sokolovski,Romano}, mixtures
of $^3$He-$^4$He with the superfluid and demixing states
\cite{Beg71,MKDietrich04,Tuyocha}, and other systems.

The properties of spin fluids were studied using mean field
(MF) theories \cite{HemImb77,TaGaTeWeNi95,Schinagl,Fefomrom03},
more accurate integral equation (IE) approaches \cite{OmMrFoFe04,%
LoWeAlBrSt94,Lado98,So98,LaLoWe98,Sokolovskii}, and Monte Carlo
(MC) simulation techniques \cite{Fefomrom03,LoWeAlBrSt94,LaLoWe98,%
Nijmeijer,Weis,Parola,Mryglod}. Different types of models, such
as the well-recognized discrete 1D spin Ising, or continuous
2D spin $XY$ and 3D Heisenberg fluids, have been considered.
Despite this, the question concerning the global phase diagram
topology of the $XY$ spin fluid including the influence of an
external magnetic field has {\em never} been addressed. Moreover,
the IE approach has been restricted either to simplified ideal
Heisenberg fluids \cite{LoWeAlBrSt94,Lado98,So98,LaLoWe98,%
Sokolovskii}, where nonmagnetic attractive interactions are
absent, or to ideal and nonideal Ising models \cite{OmMrFoFe04}.

Surprisingly, up to now there were {\em no} attempts on developing
the IE approach for the $XY$ spin fluid model. This model may play
a crucial role in the description of superfluid transitions in
pure $^4$He and its mixtures in bulk or in media such as porous
gold \cite{Tuyocha} or silica aerogel \cite{Moongir}. It is
generally believed \cite{Moongir} that the superfluid transition
in $^4$He belongs to the classical 3D $XY$ model universality
class (here 3D relates to the dimensionality of spatial
coordinates). On the other hand, the fluid of particles with
embedded $XY$ spins described by classical statistical
mechanics can be treated as one of the simplest model
of disordered continuum systems exhibiting
ferromagnetic behavior.

The presence of spin interactions and external fields destroys the
homogeneity of the fluid, producing nonuniformity or {\em anisotropy}
in the one-body density. Within the standard IE approach this
leads to the necessity of performing very complicated joint
calculations for one- and two-body distribution functions on the
basis of the coupled set of the {\em inhomogeneous} Ornstein-Zernike
(IOZ) equation, a closure relation, and the first equation
of the Born-Green-Yvon (BGY) hierarchy \cite{Hansen}. Such
calculations result in {\em unresolvable} numerical difficulties
because of the restricted capabilities of modern supercomputers.
It is worth emphasizing also that existing IE developments
for Ising \cite{OmMrFoFe04} and Heisenberg \cite{LoWeAlBrSt94,%
Lado98,So98,LaLoWe98,Sokolovskii} systems are not applicable
to the $XY$ fluid. The reason is that neither it can be mapped
onto a binary homogeneous mixture (as for Ising) nor its
anisotropic correlations be expanded in spherical harmonics
(as for Heisenberg). The specific $XY$ spin interactions
require a separate IE consideration.

In this Letter we present a method allowing to overcome the
difficulties of the IOZ approach in the case of $XY$ fluids.
Comparison of the obtained IE solutions with our simulation
results has shown a quantitative reproduction of the phase
diagrams in a wide region of temperature, density, external
field and interaction parameters.

Consider an $XY$ spin fluid model with the Hamiltonian
\begin{equation}
U\!=\! \sum_{i < j}^{N} \Big[ \phi(r_{ij}) \!-\! I(r_{ij}) \!-\!
J(r_{ij}) \, {\bf s}_i \! \cdot {\bf s}_j \Big] \!- {\bf H} \cdot \!
\sum_{i=1}^{N} {\bf s}_i \, ,
\end{equation}
where $N$ is the total number of particles, ${\bf r}_i$ is the 3D
spatial coordinate of the $i$-th body carrying 2D spin ${\bf s}_i$
of unit length, $r_{ij}=|{\bf r}_i-{\bf r}_j|$ denotes the interparticle
separation, and ${\bf H}$ is the external magnetic field vector lying
like ${\bf s}_i$ in the $XY$-plane. The exchange integral $J$ of
ferromagnetic interactions and the nonmagnetic attraction potential
$I$ can be chosen in the form of Yukawa functions,
\begin{equation}
J(r) = \frac{\epsilon \sigma}{r} \exp \! \Big(
-\frac{r-\sigma}{\sigma} \Big) \, , \ \ \ \ I(r) = J(r) \big / R \, ,
\end{equation}
where $\epsilon$ and $\sigma$ denote the interaction intensity and
the size of the particles, respectively, with $R$ being the ratio
defining the relative strength of $J$ to $I$. The repulsion $\phi$
between particles can be modeled by a more realistic soft-core
(shifted Lennard-Jones) potential
\cite{Fefomrom03,OmMrFoFe04},
\begin{equation}
\phi(r) = \left\{
\begin{array}{cc}
\displaystyle
4 \epsilon \bigg[ \Big( \frac{\sigma}{r}
\Big)^{12} - \Big( \frac{\sigma}{r} \Big)^6 \bigg] +
\epsilon \, , & \ \ r < \sqrt[6]{2} \sigma \, , \\ [12pt]
0 \, , & \ \ r \ge \sqrt[6]{2} \sigma \, ,
\end{array}
\right.
\end{equation}
rather than by the hard sphere one.

A complete thermodynamic and magnetic description of system (1)
can be performed in terms of orientationally dependent one-body
$\xi(\varphi)$ and two-body $g(r,\varphi_1,\varphi_2)=
h(r,\varphi_1,\varphi_2)+1$ distribution functions. The angles
$\varphi$ are referred to the external field, so that ${\bf H}
\cdot {\bf s} = H \cos\varphi$ and ${\bf s}_1 \! \cdot {\bf s}_2 =
\cos(\varphi_1-\varphi_2)$. According to the liquid state theory
\cite{Hansen}, the total correlation function $h$ satisfies the
IOZ equation which in our case reads

\vspace{-7pt}

\begin{eqnarray}
h(r,\varphi_1,\varphi_2) = && c(r,\varphi_1,\varphi_2) +
\frac{\rho}{2\pi} \int \limits_V {\rm d} {\bf r'} \int
\limits_{0}^{2\pi} {\rm d} \varphi
\nonumber \\ [-6pt] \\ [-6pt]
&& \times \xi(\varphi) c(|{\bf r}-{\bf r'}|,
\varphi_1,\varphi) h(r',\varphi,\varphi_2) \, ,
\ \ \ \ \nonumber
\end{eqnarray}
where $\rho=N/V$ is the particle number density, $V$ the
volume and $c(r,\varphi_1, \varphi_2)$ the direct
correlation function.

The IOZ equation (4) must be complemented by a closure relation.
The most general form of it is
\begin{equation}
g = \exp \! \big(\!-\beta u + h  - c + B \, \big) \, ,
\end{equation}
where $u(r,\varphi_1,\varphi_2)=\phi(r)-I(r)-J(r) \cos(\varphi_1-
\varphi_2)$ with $\beta^{-1}=k_{\rm B} T$ being the temperature,
and $B$ is the bridge function. This function cannot be
determined exactly for {\em any} system of interacting
particles, but a lot of approaches exist allowing to present
it approximately \cite{Hansen}. One way is to use the soft
mean spherical approximation (SMSA) \cite{OmMrFoFe04,Choudhury}
\begin{equation}
B(r,\varphi_1,\varphi_2)=\ln[1+\tau(r,\varphi_1,\varphi_2)]-
\tau(r,\varphi_1,\varphi_2) \, ,
\end{equation}
where $\tau = h - c - \beta u_{\rm l}$. The long-ranged part
$u_{\rm l}$ can be extracted \cite{OmMrFoFe04} from the total
potential $u$ as $u_{\rm l}(r,\varphi_1,\varphi_2) = - [I(r) +
J(r) \cos(\varphi_1-\varphi_2)] \exp[-\beta \phi(r)]$.

Evaluation of pair correlations from IOZ equation (4) requires the
knowledge of $\xi(\varphi)$. The latter is obtained from the first
member of the BGY hierarchy \cite{Hansen},
\begin{eqnarray}
\frac{\rm d}{{\rm d} \varphi} \ln \xi(\varphi) = &&
\frac{\rm d}{{\rm d} \varphi} \beta H \cos\varphi -
\beta \frac{\rho}{2\pi} \int \limits_V {\rm d} {\bf r} \int
\limits_{0}^{2\pi} {\rm d} \varphi'
\nonumber \\ [-5pt] \\ [-5pt] && \times
\xi(\varphi') g(r,\varphi,\varphi') \frac{{\rm d}
u(r,\varphi,\varphi')}{{\rm d} \varphi'}
\, . \nonumber \ \ \ \ \
\end{eqnarray}

Eqs.~(4), (5), and (7) constitute a very complicated set of
coupled IOZ/SMSA/BGY nonlinear integro-differential equations with
respect to $h$ (or $g$), $c$, and $\xi$. The main problem in
solving it is that the unknowns $h$ and $c$ depend on up to three
variables. This leads to unresolvable numerical difficulties, and
thus a method is needed to remedy such a situation.

Any periodic function of two angle variables can be expanded in sine
and cosine harmonics as
\begin{equation}
f(r,\varphi_1,\varphi_2) \! = \! \sum_{n,m=0}^{\infty} \,
\sum_{l,l'=0,1} \! f_{nmll'}(r) T_{nl}(\varphi_1) T_{ml'}(\varphi_2)
\end{equation}
using the orthogonal Chebyshev polynomials $T_{n0}(\varphi) =
\cos(n \varphi)$ and $T_{n1}(\varphi) = - \frac{1}{n} \, {\rm d}
T_{n0}(\varphi)/ {\rm d} \varphi =\sin(n \varphi)$. Expansion (8)
can readily be applied to our two-body functions $\{h,g,c\} \!\equiv
\! f$ with the simplification $f_{nmll'}\!=\!f_{nml} \delta_{ll'}$
because they are invariant with respect to the transformation
$(\varphi_1,\varphi_2) \leftrightarrow (-\varphi_1, -\varphi_2)$
in view of the symmetry of Hamiltonian (1). Then exploiting
the orthonormality condition $\int_{0}^{2\pi} T_{nl}(\varphi)
T_{ml'} (\varphi) {\rm d} \varphi = t_n \delta_{nm} \delta_{ll'}$,
where $t_n = \pi (1-\delta_{n0}) + 2\pi \delta_{n0}$, yields
the expansion coefficients

\vspace{-3pt}

\begin{equation}
f_{nml}(r) \!\!=\!\! \frac{1}{t_n t_m} \! \! \int \! \! \! \!
\int \! \! f(r,\varphi_1,\varphi_2) T_{nl}(\varphi_1)
T_{ml}(\varphi_2) {\rm d} \varphi_1 {\rm d} \varphi_2 .
\end{equation}

\vspace{-3pt}

In terms of these coefficients the IOZ equation (4) reduces to

\vspace{-2pt}

\begin{equation}
h_{n m l}(k) \! = c_{n m l}(k) + \rho \! \sum_{n',m'}
c_{n m' l}(k) \xi_{n' m' l} h_{n' m l}(k) \, ,
\end{equation}

\vspace{-4pt}

\noindent
where $\xi_{n m l} = \frac{1}{2\pi} \int_{0}^{2\pi} \xi(\varphi)
T_{n l}(\varphi) T_{m l}(\varphi) {\rm d} \varphi$ are the moments
of $\xi(\varphi)$, and the 3D Fourier transform $f(k)=\int_V f(r)
\exp({\rm i} {\bf k \cdot \bf r}) {\rm d} {\bf r}$ has been used.
The algebraic representation (10) looks like the OZ equation
corresponding to a mixture of ordinary {\em homogeneous} fluids.
This is a very important feature because the problem can now be
solved by adapting algorithms already {\em known} for
homogeneous systems.

Furthermore, we perform the one-body polynomial expansion

\vspace{-12pt}

\begin{equation}
\ln \xi(\varphi) = \beta H \cos\varphi + \sum_{n=0}^{\infty}
a_n T_{n0}(\varphi) \, ,
\end{equation}

\vspace{-6pt}

\noindent
where only cosine harmonics appear due to the property
$\xi(-\varphi)=\xi(\varphi)$. Then the cumbersome
integro-differential equation (7) allows to be solved
analytically,

\vspace{-5pt}

\begin{equation}
a_n = \frac{\beta \rho}{2n} \int \! {\rm d} {\bf r} \!
\mathop{\sum_{m=0}^{\infty}} \limits_{l,l'=0,1} \!
(-1)^{l+l'} \xi_{m1l} g_{n-1+2l' m l}(r) J(r)
\end{equation}

\vspace{-5pt}

\noindent
for $n \ge 1$, while the coefficient $a_0$ is determined from the
normalization $\frac{1}{2\pi} \int_{0}^{2\pi} \xi(\varphi) {\rm d}
\varphi = 1$.

Handling the SMSA closure (5) also presents no difficulties,
because for distances $r \ge 2^{1/6} \sigma$ (where $\phi(r)=0$)
we obtain from Eqs.~(5) and (6) that $c(r,\varphi_1,\varphi_2)=
\beta[I(r)+J(r) \cos(\varphi_1-\varphi_2)]$. Taking into account
the equality $\cos(\varphi_1-\varphi_2) = T_{10} (\varphi_1)
T_{10} (\varphi_2) + T_{11}(\varphi_1) T_{11} (\varphi_2)$, one
finds $c_{000}(r) = \beta I(r)$ and $c_{110}(r) = c_{111}(r) =
\beta J(r)$, while all other $c$-coefficients will be equal
to zero at $r \ge 2^{1/6} \sigma$. For $r < 2^{1/6} \sigma$,
we should perform numerical integration (see Eq.~(9)) of the
right-hand side of Eq.~(5) in order to obtain the expansion
coefficients $g_{nml}(r)$.

Another important feature is that only a {\em small} number
${\cal N}$ of harmonics should be, in fact, involved because the
expansion coefficients rapidly tend to zero with increasing
${\cal N}$. Then the sums $\sum_{n,m}^\infty$ can be replaced
without loss of precision by finite ones with $n,m \le {\cal N}$.
In our case the anisotropic potential is presented by zeroth
and first harmonics (see above the expansion for $\cos(\varphi_1-
\varphi_2)$), while a slight anharmonicity (${\cal N} > 1$) in
the correlation functions appears due to the nonlinearity
of the closure.

Once the expansion coefficients are found, {\em all} the magnetic
and thermodynamic properties of the system are obtained in a
straightforward way. In particular, the pressure $P$ can be
calculated from the virial equation

\vspace{2pt}

\begin{eqnarray}
&& \frac{\beta P}{\rho} = 1 - \frac{1}{6} \frac{\beta \rho}{(2\pi)^2}
\int \! {\rm d} {\bf r} {\rm d} \varphi_1 {\rm d} \varphi_2
\xi(\varphi_1) \xi(\varphi_2) g(r,\varphi_1,\varphi_2)
\nonumber \\ [3pt] && \times
r \frac{{\rm d} u(r,\varphi_1,\varphi_2)} {{\rm d} r}
= 1 - \frac{\beta \rho}{6} \sum_{n,m} \int \! r {\rm d} {\bf r}
\bigg( \frac{{\rm d} [\phi(r)-I(r)]}{{\rm d} r}
\nonumber \\ [-9pt] \\ [-4pt] && \times
\xi_{n00} \xi_{m00} g_{nm0}(r) - \frac{{\rm d} J(r)}{{\rm d} r}
\sum_{l=0,1} \xi_{n1l} \xi_{m1l} g_{nml}(r) \bigg) \, ,
\nonumber
\end{eqnarray}

\vspace{-3pt}

\noindent
whereas the magnetization $M\!=\!\frac{1}{2\pi} \int_{0}^{2\pi} \!
\cos(\varphi) \xi(\varphi) {\rm d} \varphi \!=\! \xi_{100}$. Then
the phase coexistence densities between gas and liquid states
can be evaluated by applying the well-known Maxwell construction
to Eq.~(13), while the P-F transition will correspond to a
boundary Curie curve in the temperature-density plane where
nonzero (spontaneous) magnetization $M \ne 0$ becomes
possible at $H=0$.

The coupled set of homogeneous OZ/SMSA/BGY equations (5), (10),
and (12) was solved by adapting the algorithm used in
Ref.~\cite{OmMrFoFe04}. The integration with respect to angle
variables has been performed by Gauss-Chebyshev quadratures.
The number of harmonics involved was ${\cal N}=3$. Further
increase of ${\cal N}$ does not affect the solutions.
Other computational details are similar to those of
Ref.~\cite{OmMrFoFe04} when solving the Ising IE equations.
The dimensionless quantities $\rho^\ast=\rho \sigma^3$,
$T^\ast=k_{\rm B} T/\epsilon$, and $H^\ast=H/ \epsilon$
were chosen in the presentation of the results.

The simulations were carried out using the Gibbs ensemble MC
(GEMC) \cite{Panagio} and multiple histogram reweighting (MHR)
\cite{Ferrenberg} techniques for evaluating the G-L and
liquid-liquid (L-L) coexistences, while the Binder crossing
scheme \cite{Mryglod,Binder} was utilized to determine the
\mbox{P-F} magnetic transition (at $H=0$). Other simulation
details are similar to those reported in Refs.~\cite{Fefomrom03,%
OmMrFoFe04,Mryglod}.

In Fig.~1 we compare the OZ/SMSA/BGY results for the {\em ideal}
($R=\infty$) $XY$ fluid at different external fields $H$ with the
GEMC and MHR simulation data. At $H=0$, a tricritical (TC) point
separates the second order P-F magnetic phase transition line from
the first order transition between a P-gas and an F-liquid. The
$H$-dependence of the G-L critical temperature and density is
{\em nonmonotonic}. Samples of the MF binodals are included in
Fig.~1 as well to demonstrate the obvious advantage of the IE
theory over the MF approach. Note that contrary to the theoretical
binodals, the GEMC and MHR coexistence curves terminate when
approaching the critical regions. This is because of the
appearance of huge density fluctuations which cannot be
properly handled within finite simulation boxes. The MHR
technique allows to approach to critical points more closely
(see Fig.~1) and should be considered as more preferable
than the GEMC method.

\vspace{1pt}

\begin{figure}[htbp]
\epsfxsize=86mm
\centerline{\epsffile{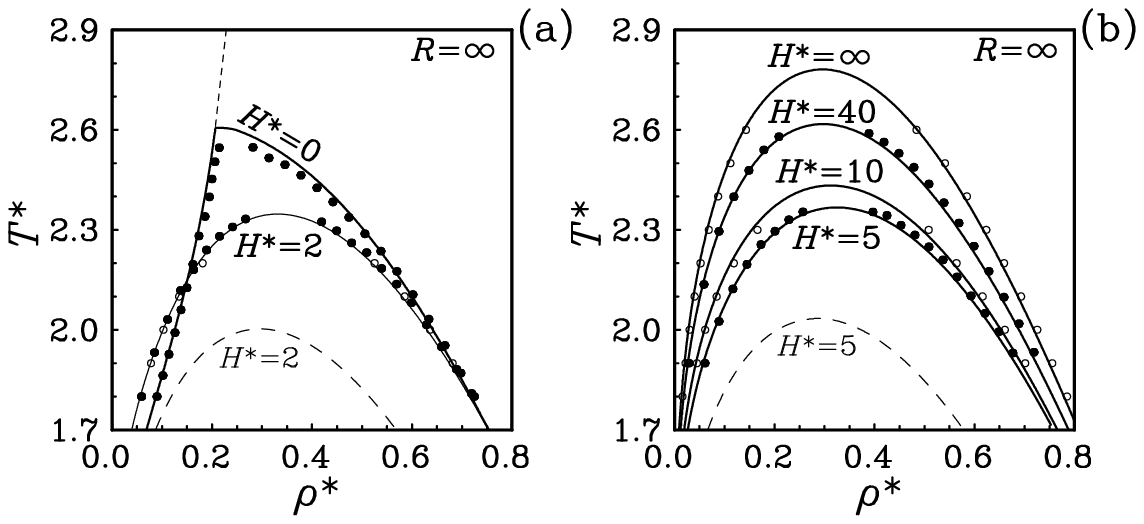}}

\addtolength{\baselineskip}{-1pt}

{\small FIG.~1. The G-L binodals obtained for the ideal $XY$ fluid
within the OZ/SMSA/BGY approach (full curves) versus the GEMC
(open circles) and MHR (full circles) simulation data. The P-F
transition is plotted by the short-dashed line. The MF samples
are shown by low lying long-dashed curves.}
\end{figure}

\vspace{3pt}

The OZ/SMSA/BGY and MHR phase diagrams of the {\em nonideal}
$XY$-fluid are shown in Figs.~2 and 3 for different ratios $R$ and
magnetic fields $H$. {\em Four} types of phase diagram topology
can be identified overall. For large $R \ge 0.415$ (type I), the
system exhibits an {\em ideal-like} behavior with the existence of
a TC point at $H=0$ and G-L transitions at $H \ne 0$ for each $R$
(Fig.~2(a)). At moderate values $0.26 < R < 0.415$ (type II), the
transition between a P-liquid and an F-liquid arises at $H=0$ {\em
additionally} to the transition between a P-gas and a P-liquid.
Here a triple point (TP) {\em occurs} too, where a rare P-gas, a
moderately dense P-liquid, and a highly dense F-liquid all coexist
at the same $T$ and $P$ (see Figs.~2(b) and 3(a)). The TPs can
{\em exist} at $H \ne 0$ as well and describe then the phase
coexistence between a weakly magnetized gas, a moderate magnetized
liquid and a strongly magnetized liquid (Fig.~3(b)). With
increasing $H$, either the G-L ($0.376 < R < 0.415$, type IIa) or
L-L ($0.26 < R < 0.376$, type IIb) transition {\em disappears} in
a critical end (CE) point at some finite $H$. For instance, even
if $R$ is slightly smaller than the boundary value $R_{\rm
vL}=0.376$, namely $R=0.37$, the L-L critical point ends at some
$H^\ast \sim 1$, while the G-L critical point extends to infinite
field (Fig.~3(a)). In the special case $R=R_{\rm vL}$, the G-L and
L-L transition lines {\em merge} into the TC van Laar point at
$H^\ast=1.9$ (Fig.~3(b)). For small $R \le 0.26$ (type III), the
translational interaction dominates over the spin one, remaining
the G-L transition, whereas the TC point at $H=0$ {\em transforms}
into a CE point (Fig.~2(b)). For $H \to \infty$, the system at any
$R$ behaves like a {\em simple} fluid with $u(r)=\phi(r)-
I(r)-J(r)$ (then all spins align along ${\bf H}$).

As can be seen, the agreement between the theory proposed and the
simulations is {\em quite} satisfactory. Slight deviations appear
only in the vicinity of critical points. This is explained by
finite size effects in the simulations and an approximate
character of the SMSA closure used in the theory. For the latter
reason, the classical value $\beta=1/2$ of the critical exponent
describing the G-L binodal behavior $|\rho-\rho_{\rm c}| \sim
|T-T_{\rm c}|^\beta$ near the criticial point ($\rho_{\rm c},
T_{\rm c}$) is recovered (in particular, at $R=\infty$ and $H \ne
0$), instead of the value $\beta \approx 1/3$ known from the
renormalization group analysis \cite{Zinj}. On the other hand, the
crossover to the TC value $\beta=1/4$ can be observed near the van
Laar point at $R=0.376$ and $H^\ast=1.9$ (Fig.~3(b)).

\vspace{1pt}

\begin{figure}[htbp]
\epsfxsize=86mm
\centerline{\epsffile{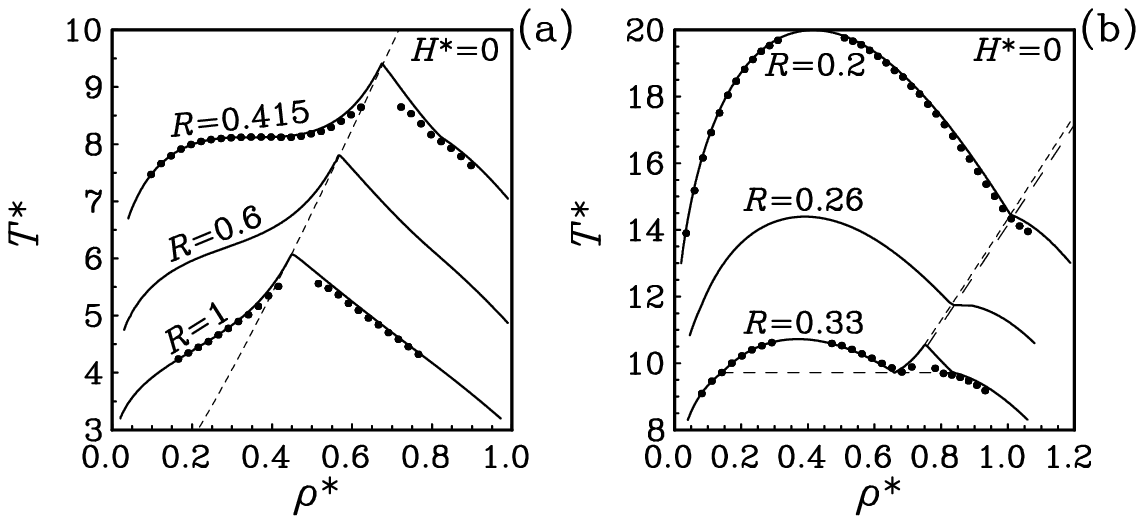}}

\addtolength{\baselineskip}{-1pt}

{\small FIG.~2. The G-L and L-L binodals of the nonideal $XY$
fluid within the OZ/SMSA/BGY approach (full curves) versus the MHR
data (circles). The P-F transition is plotted by the short-
(theory) and long- (simulation) dashed curves. The triple point
is represented by the horizontal dashed line.}
\end{figure}

\vspace{-10pt}

\begin{figure}[htbp]
\epsfxsize=86mm
\centerline{\epsffile{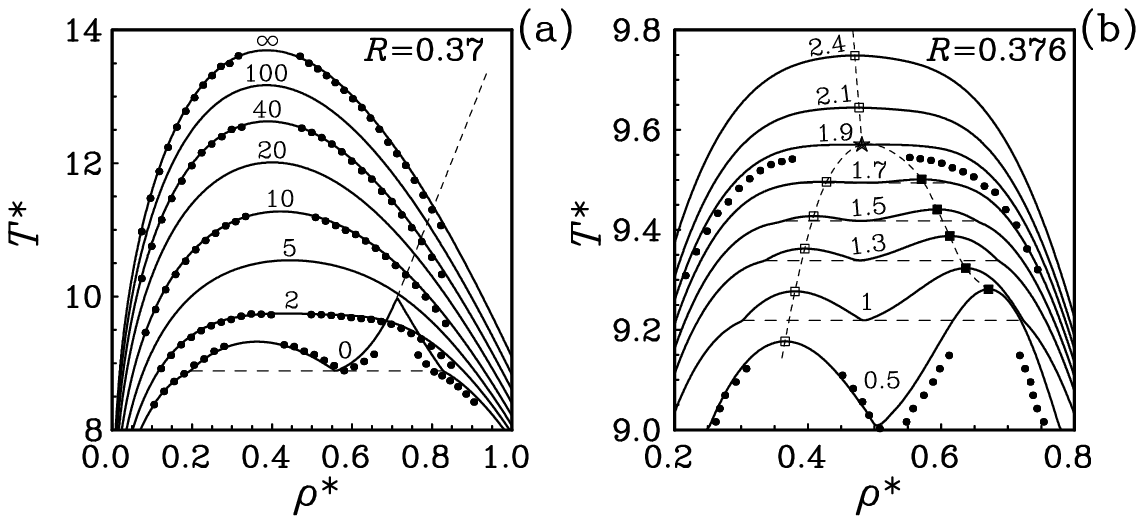}}

\addtolength{\baselineskip}{-1pt}

{\small FIG.~3. The binodals near (a) and at (b) the boundary
value $R=R_{\rm vL}$. The G-L and L-L critical points are shown in
subset (b) for different $H^\ast$ as open and full squares,
respectively, connected by dashed curves. The curves meet in the
TC point (star). Other notations are the same as for Fig.~2.}
\end{figure}

\vspace{1pt}

More precise IE calculations near critical points are {\em possible}
provided a more accurate closure is used. For instance, the
self-consistent OZ ansatz (SCOZA) \cite{Kahlang,Kahl,Paschinger}
(which in its present formulation was implemented only for simple
homogeneous hard-sphere Yukawa systems) can be {\em extended} to
our inhomogeneous soft-core $XY$ fluid by introducing a state
dependent function $K(\rho,T,H)$ into the SMSA closure. Then
$K$ is determined by the requirement of thermodynamic {\em
consistency} between the energy and compressibility routes.
In view of the inhomogeneity and softness, this leads to
a significant sophistication of the calculations. They
go beyond the scope of the present Letter and will be
considered elsewhere.

In conclusion, we point out that a {\em novel} technique to study
orientationally ordered fluids with planar spins has been
proposed. It combines the standard IE method with appropriate
expansions of the inhomogeneous correlation functions in terms of
orthogonal polynomials. This reduces the calculations to those
inherent in a mixture of ordinary homogeneous fluids and thus
presents now {\em no} numerical difficulties. Detailed comparisons
with simulations have shown that the proposed approach is powerful
enough to give a {\em quantitative} description of phase
transitions in the $XY$ spin fluid systems.

\vspace{2pt}

This work was supported in part by the Fonds zur F\"orderung der
wissenschaftlichen Forschung under Project No.~P15247.

\end{multicols}


\vspace{-14pt}

\begin{thebibliography}{99}

\vspace*{-48pt}

\bibitem{HemImb77}
 P. C. Hemmer and D. Imbro, Phys. Rev. A {\bf 16}, 380 (1977).

\bibitem{TaGaTeWeNi95}
 J. M. Tavares {\em et al.},
 Phys. Rev. E {\bf 52}, 1915 (1995).

\bibitem{Schinagl}
 F. Schinagl, H. Iro, and R. Folk, Eur. Phys. J. B {\bf 8}, 113 (1999).

\bibitem{Fefomrom03}
 W. Fenz {\em et al.},
 Phys. Rev. E {\bf 68}, 061510 (2003).

\bibitem{OmMrFoFe04}
 I. P. Omelyan {\em et al.},
 Phys. Rev. E {\bf 69}, 061506 (2004).

\bibitem{Wilding}
 N. B. Wilding, F. Schmid, and P. Nielaba, Phys. Rev. E
 {\bf 58}, 2201 (1998).

\bibitem{Kahlang}
 G. Kahl, E. Sch\"oll-Paschinger, and A. Lang, Monatshefte
 f\"ur Chemie {\bf 132}, 1413 (2001).

\bibitem{Kahl}
 G. Kahl, E. Sch\"oll-Paschinger, and G. Stell, J. Phys.:
 Condens. Matter {\bf 14}, 9153 (2002).

\bibitem{Paschinger}
 E. Sch\"oll-Paschinger and G. Kahl, J. Chem. Phys. {\bf 118},
 7414 (2003).

\bibitem{Pini}
 D. Pini {\em et al.},
 Phys. Rev. E  {\bf 67}, 046116 (2003).

\bibitem{Sokolovski}
 R. O. Sokolovskii, Phys. Rev. B {\bf 61}, 36 (2000).

\bibitem{Romano}
 S. Romano and R. O. Sokolovskii, Phys. Rev. B {\bf 61}, 11379 (2000).

\bibitem{Beg71}
 M. Blume, V. J. Emery, and R. B. Griffiths, Phys. Rev. A {\bf 4}, 1071 (1971).

\bibitem{MKDietrich04}
 A. Maciolek, M. Krech and S. Dietrich, Phys. Rev. E {\bf 69}, 036117 (2004).

\bibitem{Tuyocha}
 D. J. Tulimieri, J. Yoon, and M. H. W. Chan,
 Phys. Rev. Lett. {\bf 82}, 121 (1999).

\bibitem{LoWeAlBrSt94}
 E. Lomba {\em et al.},
 Phys. Rev. E {\bf 49}, 5169 (1994).

\bibitem{Lado98}
 F. Lado and E. Lomba, Phys. Rev. Lett. {\bf 80}, 3535 (1998).

\bibitem{So98}
 T. G. Sokolovska, Physica A {\bf 253}, 459 (1998).

\bibitem{LaLoWe98}
 F. Lado, E. Lomba, and J. J. Weis, Phys. Rev. E {\bf 58}, 3478 (1998).

\bibitem{Sokolovskii}
 T. G. Sokolovska and R. O. Sokolovskii, Phys. Rev. E {\bf 59}, R3819 (1999).

\bibitem{Nijmeijer}
 M. J. P. Nijmeijer and J. J. Weis, Phys. Rev. E {\bf 53}, 591 (1996).

\bibitem{Weis}
 J. J. Weis {\em et al.},
 Phys. Rev. E {\bf 55}, 436 (1997).

\bibitem{Parola}
 M. J. P. Nijmeijer, A. Parola, and L. Reatto, Phys. Rev. E
 {\bf 57}, 465 (1998).

\bibitem{Mryglod}
 I. M. Mryglod, I. P. Omelyan, and R. Folk,
 Phys. Rev. Lett. {\bf 86}, 3156 (2001).

\bibitem{Moongir}
 K. Moon and S. M. Girvin, Phys. Rev. Lett. {\bf 75}, 1328 (1995).

\bibitem{Hansen}
 J. P. Hansen and I. R. McDonald, {\em Theory of Simple Liquids}, 2nd
 edn. (Academic, London, 1986).

\bibitem{Choudhury}
 N. Choudhury and S. K. Ghosh, J. Chem. Phys. {\bf 116}, 8517 (2002).

\bibitem{Panagio}
 A. Z. Panagiotopoulos, Molec. Sim. {\bf 9}, 1 (1992).

\bibitem{Ferrenberg}
 A. M. Ferrenberg and R. H. Swendsen, Phys. Rev. Lett. {\bf 61},
 2635 (1988); {\bf 63}, 1195 (1989).

\bibitem{Binder}
 K. Binder, Rep. Prog. Phys. {\bf 60}, 487 (1997).

\bibitem{Zinj}
 J. Zinn-Justin, {\em Quantum Field Theory and Critical Phenomena}
 (Clarendon, Oxford, 1983).

\end{thebibliography}
\end{document}